\documentclass[12pt,authoryear]{article}                     
\usepackage{mathptmx}
\usepackage{graphicx}
\usepackage[utf8x]{inputenc}
\usepackage{subfigure}
\usepackage{latexsym}
\usepackage{amsmath}
\usepackage{amsfonts}
\usepackage{color,amssymb}
\usepackage{natbib}
\usepackage{hyperref}
\usepackage[affil-it]{authblk}

\usepackage[affil-it]{authblk}

\begin{document}
\title{Fast estimation of posterior probabilities in change-point analysis through a constrained hidden Markov model}

\author{The Minh Luong%
  \thanks{e-mail: \texttt{the-minh.luong@parisdescartes.fr}; Corresponding author}}

\author{Yves Rozenholc%
}

\author{Gregory Nuel%
}
\affil{MAP5, Universit\'{e} Paris Descartes, 45 rue des Saints-P\`{e}res, 75006 Paris, France}

\date{July 2012}
\maketitle

\begin{abstract}
The detection of change-points in heterogeneous sequences is a statistical challenge with applications across a wide variety of fields. In bioinformatics, a vast amount of methodology exists to identify an ideal set of change-points for detecting Copy Number Variation (CNV). While considerable efficient algorithms are currently available for finding the best segmentation of the data in CNV, relatively few approaches consider the important problem of assessing the uncertainty of the change-point location. Asymptotic and stochastic approaches exist but often require additional model assumptions to speed up the computations, while exact methods have quadratic complexity which usually are intractable for large datasets of tens of thousands points or more. In this paper, we suggest an exact method for obtaining the posterior distribution of change-points with linear complexity, based on a constrained hidden Markov model. The methods are implemented in the R package \textit{postCP}, which uses the results of a given change-point detection algorithm to estimate the probability that each observation is a change-point. We present the results of the package on a publicly available CNV data set $(n=120)$. Due to its frequentist framework, postCP obtains less conservative confidence intervals than previously published Bayesian methods, but with linear complexity instead of quadratic. Simulations showed that $\textit{postCP}$ provided comparable loss to a Bayesian MCMC method when estimating posterior means, specifically when assessing larger-scale changes, while being more computationally efficient. On another high-resolution CNV data set $(n=14,241)$, the implementation processed information in less than one second on a mid-range laptop computer. 

\end{abstract}


\section{Introduction}
The detection of \emph{change-points} in heterogeneous sequences is a statistical challenge with many applications in fields such as finance, reliability, signal analysis, neurosciences and biology \citep{pinkel98,snijders01}. In bioinformatics in particular, a vast amount of methodology \citep{olshen04,fridlyand04,hupe04} exists for identifying an ideal set of change-points in data from array Comparative Genomic Hybridization (aCGH) techniques, in order to identify Copy Number Variation (CNV). 

A typical expression of the change-point problem is as follows, given data $X=(X_1,X_2,\ldots,X_n)$ of real-valued observations, $(S_1,\ldots,S_n)$ corresponding segment indices of the observations, and $\mathcal{M}_K$ as the set of all possible combinations of $S$ for fixed $K\geqslant 2$ number of segments. The goal is to find the best partitioning $S \in \mathcal{M}_K$  into $K$ non-overlapping intervals, assuming that the distribution is homogeneous within each of these intervals. 

For $K$ segments of contiguous observations, the \emph{segment-based model} expresses the distribution of $X$ given a segmentation $S \in \mathcal{M}_K$ as: 
\begin{equation}\label{eq:themodel}
\mathbb{P}(X | S; \theta)=\prod_{i=1}^{n} g_{\theta_{S_i}}\left(X_i\right) = \prod _{k=1}^K \prod_{i,S_i=k}  g_{\theta_k}\left(X_i\right)
\end{equation}
where $g_{\theta_k}(\cdot)$ is the parametric, or emission, distribution (e.g.: normal or Poisson) of the observed data with parameter $\theta_k$, $\theta=(\theta_1,\ldots,\theta_K)$ is the global parameter, and $S_i$ is the segment index at position $i$. For example, if $n=5$, $K=2$, and with change-point located between positions $2$ and $3$, then $S=(1,1,2,2,2)$.

Introducing a prior distribution $\mathbb{P}(S)$ on any $S \in \mathcal{M}_K$ obtains a posterior distribution of the segmentation:
\begin{equation}
\mathbb{P}(S|X;\theta)=\frac{\mathbb{P}(X|S;\theta)\mathbb{P}(S)}{\sum_R \mathbb{P}(X|R;\theta)\mathbb{P}(R)}.
\end{equation}

For a uniform prior, set $\frac{1}{\mathbb{P}(S)}={n-1 \choose K-1}=|\mathcal{M}_K|$.

A common alternative to the above segmentation procedure is to consider an unsupervised hidden Markov model (HMM). Assuming that $S$ is a Markov chain of hidden states, this approach \citep{rabiner89} can be thought of as being \emph{level-based}, with the parameter of the $k^{th}$ segment takes its value in the set of $L\geqslant 1$ levels~\footnote{Similar to the segment-based model, the choice of $L$ is critical and is usually addressed through penalized criteria.}: $\{\theta_1,\theta_2,\ldots,\theta_L\}$. This simply is equivalent to the model defined by Equation~(\ref{eq:themodel}), with the noticeable difference that $S \in \{1,2,\ldots,L\}^n$. With this level-based approach $K\geqslant L$ in general, and the HMM is unconstrained in the sense that transitions are possible between any pair of states. This is an appropriate model when the conditional distribution within a given segment of contiguous observations may be shared among other segments. While the unconstrained HMM is preferable in many practical situations, the segment-based model requires less assumptions and is thus a more general model. 

A convenient feature of these HMM approaches is in computing efficiently the posterior distribution $\mathbb{P}(S|X;\theta)$ in $O(L^ 2 n)$ using classical forward-backward recursions \citep{durbin98}, making them suitable for handling large datasets. This paper focuses on using an computationally efficient exact procedure to characterize the uncertainty $\mathbb{P}(S|X;\theta)$ of the estimated change-point locations using a hidden Markov model adapted to the conventional segmentation model as previously described. We exploit the effectiveness of the level-based HMM approach through a constrained HMM corresponding \emph{exactly} to the above segment-based model, providing a fast algorithm for computing $\mathbb{P}(S|X;\theta)$.  

We develop this posterior distribution procedure as change-point assessment in practical applications becomes more challenging from a computational point of view. For example, emerging high-throughput technologies are producing increasingly large amounts of data for CNV detection. For finding the exact posterior distribution of change-points $\mathbb{P}(S|X;\theta)$, \citet{guedon08} suggested an algorithm in $O(K n^2)$, while \citet{rigaill11} considered the same quantity in a Bayesian framework with the same complexity. However, the complexity of these approaches provides for very slow processing for large datasets with sequences of tens of thousands or more. Other estimates generally focus on asymptotic behavior whose conditions are delicate due to the discreteness of the problem \citep{bai03,muggeo03}, on bootstrap techniques \citep{huskova08} and on stochastic methods such as particle filtering \citep{fearnhead03}, recursive sampling \citep{lai08}, and Markov chain Monte Carlo \citep{erdman08}. Furthermore, many of the faster stochastic algorithms assume normal error structure to speed up the estimation procedures and are thus more difficult to adapt to non-normal data \citep{lai08,erdman08}.

Section~\ref{methods} presents a summary of current change-point methods, the constrained HMM algorithm and a description of the accompanying \textit{R} statistical package, Section~\ref{aCGH} implements the methods on a published array CGH data set and compares with published results, Section~\ref{examples} presents examples of simple simulated data sets with comparisons between methods, Section~\ref{CNV} shows an illustrative example of the methods on a larger scale SNP array data set, while Section~\ref{discussion} includes a discussion.

\section{Methods\label{methods}}

\subsection{Current change-point detection methods}

A considerable amount of literature focuses on detecting the ideal number or set of change-point locations. For CNV detection, \citet{fridlyand04} developed a discrete HMM to map the number of states and the most likely state at each position. Observations where state transitions are most likely to occur indicate change-points. Various extensions to this HMM approach include various procedures such as merging change-points \citep{willenbrock05} and specifying prior transition matrices \citep{marioni06} to improve the results, and simultaneous identification of CNV across multiple samples \citep{shah09}. HMM-based implementations for dealing with higher-resolution data for current array technologies include those from \citet{colella07,wang07}. 

A wide amount of segment-based approaches for identifying CNV in genomics data have also been explored. \citet{olshen04} introduced an extension of binary segmentation through a non-parametric approach, using permutation rather than parameter estimation to test for change-points. This algorithm requires data smoothing and pruning to improve computation time and change-point detection for very large sequences. To speed up this computationally-intensive process, \citet{venkatraman07} introduced adjusted p-values and stopping rules for the permutations. \citet{hupe04} introduced a likelihood-based approach to estimate parameters in a normal Gaussian model after adaptive weights smoothing. \citet{hsu05} proposed a wavelet-based change-point detection procedure with data smoothing, while \citet{eilers05} introduced a quantile approach to smoothing. \citet{willenbrock05,lai05} summarized and compared the various segmentation approaches for aCGH data.

To estimate the number of segments, $K$, \citet{zhang07} extended an earlier method including a modified Bayes Information Criterion to adjust for the number of change-points using a recursive procedure. \citet{comte04} described a least squares minimization procedure with a penalization criterion for the number of contrasts, while \citet{picard05} implemented an adaptive method for estimating the location and number of change-points, with penalization terms for likelihoods.

\subsection{Constrained HMM}

Let us assume that $S$ is a heterogeneous Markov chain over $\{1,2,\ldots,K,K+1\}$ (where $K+1$ is a ``junk'' state only considered for consistency reasons) such as $\mathbb{P}(S_1=1)=1$ and with the following transitions: for all $2 \leqslant i \leqslant n$, and $1 \leqslant k \leqslant K$ we have $\mathbb{P}(S_i =k | S_{i-1}= k)= 1-\eta_{k}(i)$ and $\mathbb{P}(S_i =k+1 | S_{i-1}= k)= \eta_{k}(i)$. For consistency, choose $\mathbb{P}(S_i=K+1|S_{i-1}=K+1)=1$. For example, if $n=5$, $K=2$, and $S=(1,1,2,2,2)$ then $\mathbb{P}(S)=(1-\eta_1(2))\eta_2(3)(1-\eta_2(4))(1-\eta_2(5))$. With this Markov chain, it is clear that $\{S \in \mathcal{M}_K\}=\{S_n=K\}$. 

In the particular case where the Markov chain is homogeneous with $\eta_k(i)=\eta \in ]0,1[$ for all $k$ and $i$, $\mathbb{P}(S)=(1-\eta)^{n-K}\eta^{K-1}$ for all $S \in \mathcal{M}_K$ ~\footnote{Note that the particular value of $\eta$ does affect $\mathbb{P}(S)$ but has no effect whatsoever on $\mathbb{P}(S|S\in \mathcal{M}_K)$ of the chosen segmentation $S$. We can therefore safely make an arbitrary choice like $\eta=0.5$ for practical computations.}. In other words, only positive state jumps of $+1$ are possible. Therefore $\mathbb{P}(S| S \in \mathcal{M}_K)=1/|\mathcal{M}_K|$, which corresponds to the canonical choice of a uniform prior on $\mathcal{M}_K$. Note that a choice of different transition coefficients $\eta_{k}(i)$ allows us to specify informative priors.

\subsection{Forward-backward procedure and posterior probabilities\label{forwback}}

The constrained HMM provides a framework for additional inference on the uncertainty in the estimated change-point model; in particular, after obtaining the segmentation from any previous procedure, we can obtain confidence intervals around each of the identified change-points. In terms of practical applications, this approach is helpful when dealing with situations where both very short and very long segments may be present, and the exact location of change-points may not be identifiable. 

The forward-backward algorithm \citep{rabiner89}, also known as posterior encoding, is a recursive algorithm that can estimate the posterior probabilities of each observation $i$ being in a particular hidden state $S_i$, and being a change-point such that $S_i\neq S_{i-1}$. The algorithm is of complexity $O(K n)$ for $K$ segments and $n$ observations; the sparse transition matrix between states reduces the $O(K^2 n)$ complexity of the classical forward-backward algorithm. A summary of this algorithm and other inferential procedures involved in HMM estimation is in \citet{cappe05}.

We define the forward and backward quantities as follows, for observation $i$ and state $k$:

For $1 \leqslant i \leqslant n-1$:
\begin{align}
F_i(k)&=\mathbb{P}(X_{1:i}=x_{1:i},S_i=k)\\
B_i(k)&=\mathbb{P}(X_{i+1:n}=x_{i+1:n},S_n=K|S_i=k)
\end{align}

We obtain the forward quantities by recursion through the following formulae:

\noindent Forward:
\begin{align}
F_1(k)&=\left\{
\begin{array}{ll}
g_{\theta_1}(x_1) & \text{if $k=1$}\\
0 & \text{else}
\end{array}
\right.\\
\nonumber F_i(k)&=\sum_{\ell} F_{i-1}(\ell)\mathbb{P}(S_i=k | S_{i-1}=\ell,S\in \mathcal{M}_K )g_{\theta_k}(x_i)\\
&=[F_{i-1}(k)(1-\eta_k(i))+\mathbf{1}_{k>1} F_{i-1}(k-1) \eta_k(i) ]g_{\theta_k}(x_i)
\end{align}

where $g_{\theta_k}(x_i)$ is the density function of the chosen emission distribution $g$ with parameter $\theta_k$, when $k$ is the underlying segment for observation $i$.

We use a similar recursive procedure to obtain the backward quantities:

\noindent Backward:
\begin{align}
B_{n-1}(k)&=\left\{
\begin{array}{ll}
\eta_K(n) g_{\theta_k}(x_n)& \text{if $k=K-1$}\\
(1-\eta_K(n))g_{\theta_k}(x_n) & \text{if $k=K$}\\
0 & \text{else}
\end{array}
\right.\\
\nonumber B_{i-1}(k)&=\sum_{\ell} \mathbb{P}(S_i=\ell | S_{i-1}=k,S\in \mathcal{M}_K)g_{\theta_\ell}(x_i)B_{i}(\ell)\\
&=(1-\eta_k(i))g_{\theta_k}(x_i)B_{i}(k) +\mathbf{1}_{k<K} \eta_{k+1}(i) g_{\theta_{k+1}}(x_i)B_{i}(k+1)
\end{align}

To obtain the posterior probabilities of the state $S_i=k$ at position $i$, we note that :
\begin{align}
&\mathbb{P}(X_{1:n}=x_{1:n},S\in \mathcal{M}_K)&=F_1(1)B_1(1)\\
&\mathbb{P}(S_i=k|X_{1:n}=x_{1:n},S\in \mathcal{M}_K)&=\frac{F_{i}(k) B_{i}(k)}{F_1(1)B_1(1)}.
\end{align}

The constrained HMM estimates the probability of changing state while being at state $S_i=k$ at observation $i$ as:
\begin{align}
\mathbb{P}(S_{i}=k|X_{1:n}=x_{1:n},S_{i-1}=k-1,S\in \mathcal{M}_K)=\frac{B_{i}(k)\eta_{k-1}(i)g_{\theta_{k}}(x_{i})}{B_{i-1}(k-1)}.\label{cp_eqn}
\end{align}
We can sample a vector of length $K-1$ change-points from the original data set through these quantities, and afterwards generate a new vector of observed data through the chosen emission distributions.
The posterior probability of the $k^{th}$ change-point occurring after observation $i$, or in other words $i+1$ being the first observation in the $k+1^{th}$ segment, is:
\begin{align}
\nonumber \mathbb{P}(CP_k=i&|X_{1:n}=x_{1:n},S\in \mathcal{M}_K)=\mathbb{P}(S_i=k,S_{i+1}=k+1|X_{1:n}=x_{1:n},S\in \mathcal{M}_K)\\
&=\frac{F_i(k) \eta_{k}(i+1) g_{\theta_{k+1}}(x_{k+1})B_{i+1}(k+1)}{F_1(1)B_1(1)} 
\end{align}

The accuracy of these posterior probabilities relies on the ability of the preceding change-point model to provide proper initial estimates of the number of and locations of the change-points, as they are crucial in selecting the $\theta_k$ parameters used in the forward-backward procedures.

Maximum \textit{a posteriori} estimation of the most probable set of change-points is also possible through the Viterbi algorithm \citep{viterbi67} by modifying the forward quantities, with the Viterbi quantities chosen as $V_i(k)$ for observation $i$ and state $k$.

\begin{align*}
V_{1}(1)&=g_{\theta_1}(x_1),\\
V_{i}(1)&=V_{i-1}(1)(1-\eta_1(i))g_{\theta_1}(x_i), \text{ if $i\geq 2,k=1$}\\
V_{i}(k)&=\max\left\{V_{i-1}(k-1)\eta_k(i),V_{i-1}(k)(1-\eta_k(i))\right\}g_{\theta_k}(x_i), \text{ if $i,k\geq 2$}
\end{align*}

We trace the path of indices $k$ used to calculate $V_{i,k}$ to get the set of change-points with highest posterior probability:
\begin{itemize}
\item $K-1^{th}$ change-point $CP_{K-1}$: $i$ of $V_{i}(K-1)$ used to calculate $V_{i+1}(K)$
\item $k^{th}$ change-point $CP_{k}$: $i$ of $V_{i}(k)$ used to calculate $V_{i+1}(k+1)$ (where $CP_{k}<CP_{k+1}$).
\end{itemize}

\subsection{Statistics package postCP}

We apply the preceding methods in the statistics package postCP, available on the CRAN website \\\url{http://cran.r-project.org/web/packages/postCP}.\\ The forward and backward recursive algorithms are programmed in C++ to optimize the speed of the computationally-intensive process.

The following is a typical $R$ command line for segmenting a sequence $LRR$, with a vector of length $K-1$ change-points (or last index of segments) $initseg$, and $95\%$ confidence intervals. The options also save forward-backward and posterior change-point probabilities in the output, obtain the most likely set of change-points through the Viterbi algorithm, and generate $100$ different vectors (each of length $K-1$) of change-point locations for generating data according to our change-point model. 
\begin{verbatim}
postCP(data=LRR,seg=initseg,ci=0.95,viterbi=TRUE,nsamples=100)
\end{verbatim}
The package also provides documentation for options on specifying a matrix of log-densities corresponding to any specified distribution of the data, and the prior specification of a matrix of transition probabilities. 

\section{Detecting copy number variation in array CGH breast cancer data\label{aCGH}}

We apply the methods to a widely referenced data set from cell line BT474 from a breast cancer tumor \citep{snijders01}. The data consist of log-reference ratios signifying the ratio of genomic copies of test samples compared to normal. The goal is to segment the data into segments with similar copy numbers, with change-points corresponding to copy number aberrations pointing to possible genes of interest \citep{pinkel98}. We compare the results of postCP to the Bayesian confidence intervals previously published on the same data by \citet{rigaill11}, consisting of $120$ observations from chromosome~10. The observations are sorted according to their relative position along chromosome~10.

We use a modification of the greedy least squares $K$-means algorithm \citep{hartigan79} to obtain an initial segmentation for $3$ and $4$ segments (Table~\ref{seg_acgh}). The locations refer to the indices of the last observations of the first $K-1$ segments. Afterwards, we used \textit{postCP} to obtain estimates of the forward and backward matrices in Section~\ref{forwback}, and afterwards, estimates of the posterior probabilities of the change-points being at each observation. We assumed a homoscedastic normal model for the observations. 

\begin{figure*}
\begin{center}
\subfigure[Three segments]{
\includegraphics[scale=0.35]{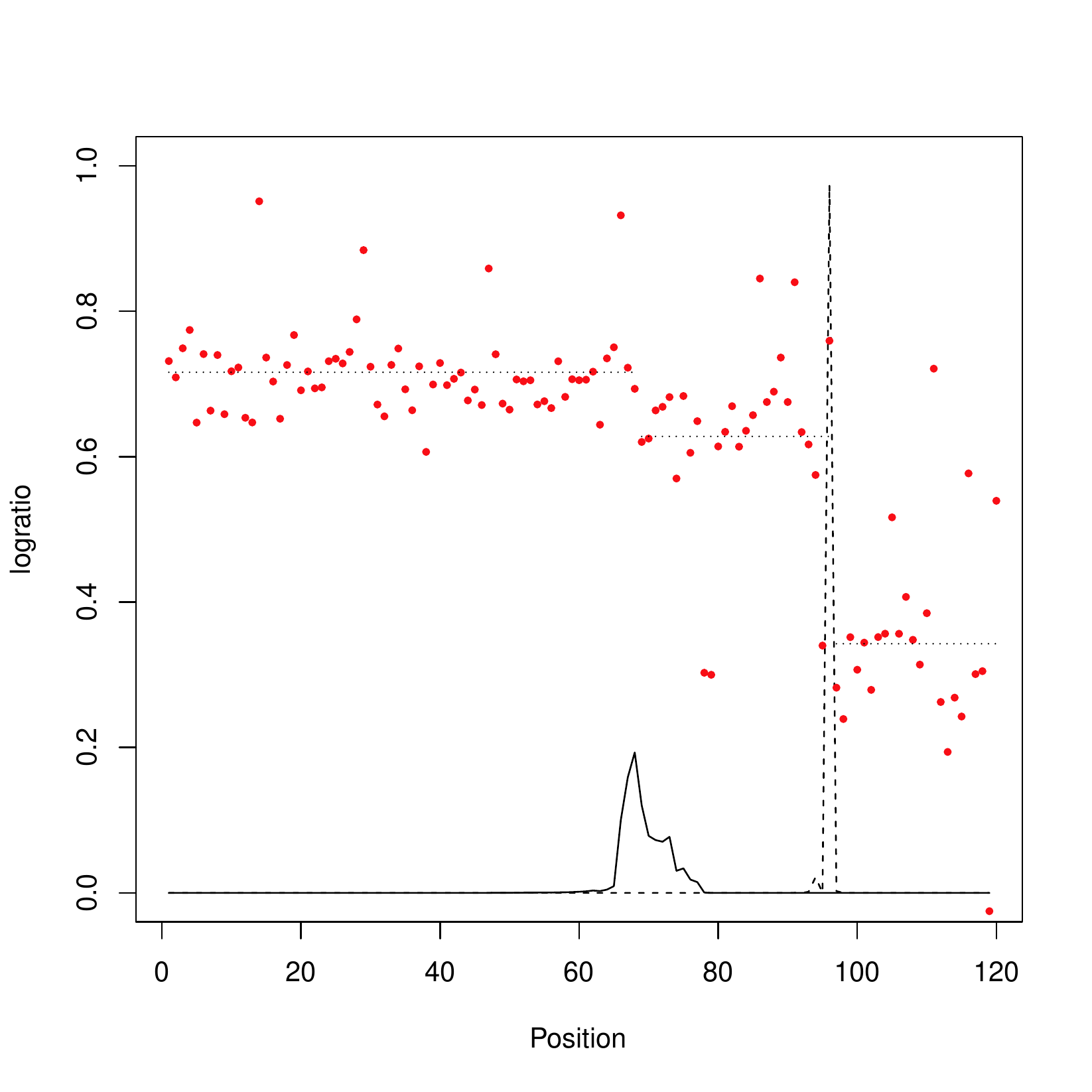}
\label{acgh_3seg}}       
\subfigure[Four segments]{
\includegraphics[scale=0.35]{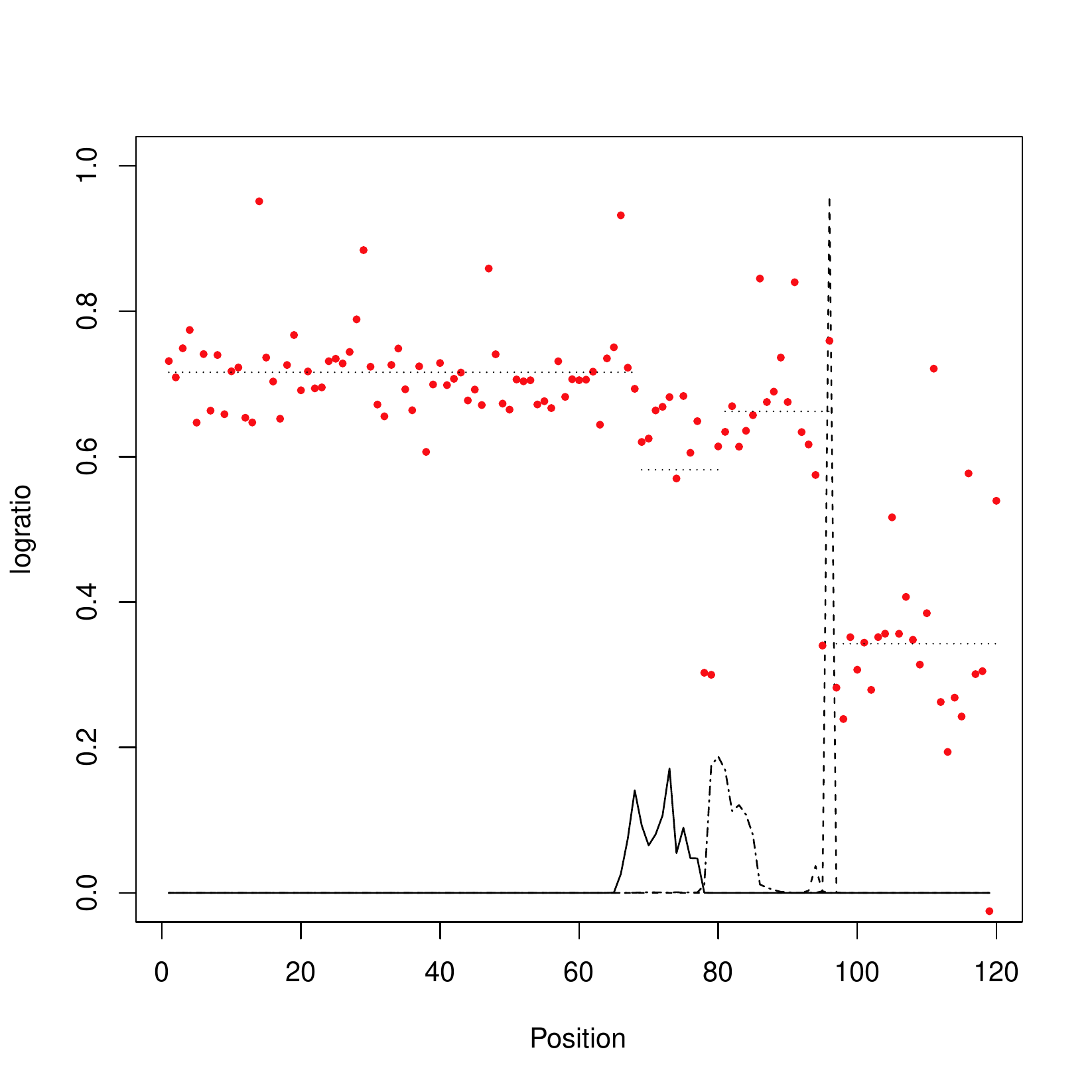}
\label{acgh_4seg}}    
\caption{Plots of estimated posterior change-point probabilities for chromosome~10 data \citep{snijders01}. Dots are LRR data with horizontal lines representing the estimated means within segments scaled to the probability axes. (a) Posterior probabilities for three segments, first change-point initialized after $i=68$ is a solid line, 2nd initialized after $i=96$ is a dashed line. (b) Posterior probabilities for four segments, additional change-point after $i=80$ is a dotted-dashed line.}
\end{center}
\end{figure*}

Figure~\ref{acgh_3seg} displays the estimated posterior change-point probabilities of the aCGH data for three segments and Figure~\ref{acgh_4seg} for four segments. For both segmentations, the probability of the last change-point had a sharp peak close to 1.0. There is an additional change-point found at position 80 by the four segment model with a relatively high uncertainty. In both the 3- and 4-segment instances, the shapes of the change-point distributions are irregular due to the discreteness of the segmentation procedure.

At these predefined observations, the corresponding posterior change-point probabilities were higher than reported by Rigaill, and confidence intervals were slightly narrower. This is expected, as Rigaill's method uses a Bayesian framework that accounts for the uncertainty in the parameter estimates. In particular, the rightmost change-point estimation with both $3$ and $4$ segments had posterior probabilities above $0.95$, while the corresponding posterior probabilities from the Bayesian method were closer to 0.5. 
\begin{table}
\begin{center}
\caption{\label{seg_acgh} Change-point confidence intervals aCGH chromosome 10} 
\setlength{\tabcolsep}{1mm}
\begin{tabular}{lrrrr}\hline\hline\noalign{\smallskip}
CP \#&$\Delta$ mean&Est location&95 \% CI (postCP)&$95\%$ CI (Bayesian)\\\hline
  \multicolumn{5}{c}{Three segments}\\
 1& -0.22  &68& 66-76 & 64-78\\
 2& -0.71  &96& 96-96 & 92-97\\
 \multicolumn{5}{c}{Four segments}\\
 1& -0.34  &68& 66-76 & 66-78\\
 2& -0.20 &80& 79-85 & 78-97\\
 3& -0.80  &96& 96-96 & 91-112\\ \hline
\end{tabular}
\flushleft{ Change-point estimates of aCGH data for chromosome~10 from \citet{snijders01}, with (95\% confidence interval), by \textit{postCP} and Bayesian confidence intervals by \citet{rigaill11}. Narrower confidence intervals were found by postCP.}
\end{center}
\end{table}

\begin{figure*}
\begin{center}
  \includegraphics[scale=0.5]{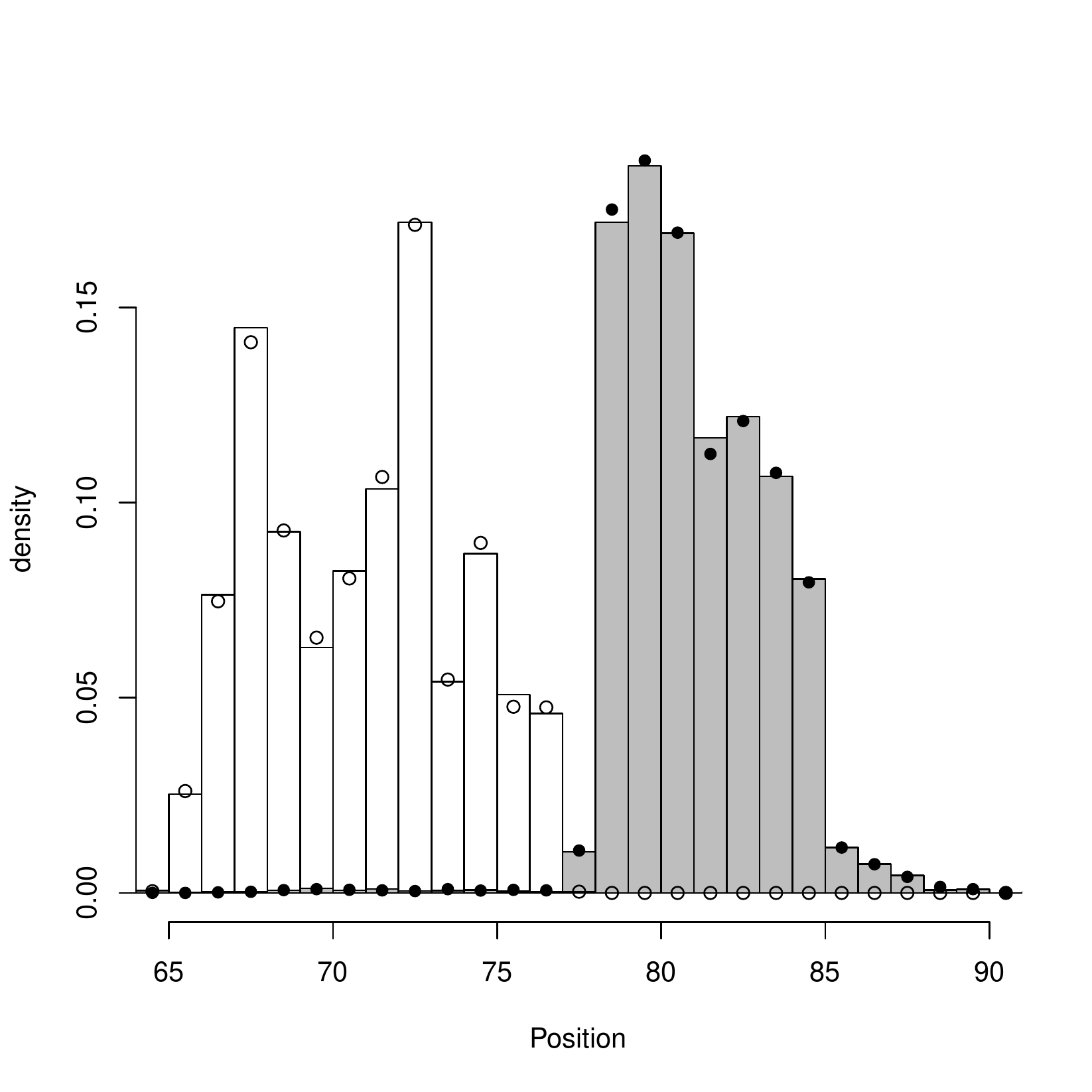}
\caption{Histogram of locations of the first two change-points, randomly and simultaneously generated by \textit{postCP} for 10,000 sets, for data from \citet{snijders01}. The frequencies of the first change-point from the simulations are in white boxes, of the second in gray boxes. The estimated change-point probabilities by the forward-backward algorithm for the first change-point are in white dots, for the second change-point in black. The randomly generated change-points are close to those estimated by postCP.}
\label{aCGH_hist}       
\end{center}
\end{figure*}

We also can use \textit{postCP} to obtain a joint sample of an entire set of change-points from the original data using the constrained HMM model. Using postCP, we sampled $10,000$ data sets using (\ref{cp_eqn}). As seen in Figure~\ref{aCGH_hist}, the histogram of the generated locations for the first two change-points closely approximates the posterior change-point probabilities found by postCP. Though there is little overlap between the estimated confidence intervals for the first two change-points, there was still a small but non-zero empirical Pearson correlation $(r=0.123,p<0.001)$ between the simultaneously generated first and second change-point locations. On the other hand, there was no association between the third generated change-point and the other two change-points, an expected result since their confidence intervals do not overlap.
\section{Change-point data simulations and comparisons \label{examples}}

In this section we consider different change-point situations to compare the ability of the methods in assessing uncertainty. In this section, we use a loss function to assess the ability of the change-point methods in estimating the posterior mean of each observation $(i=1,\ldots,n)$. For normally distributed data, we quantify loss through the mean square error (MSE). We define $MSE=\underset{i}{\sum}(\hat{\theta}(i)-\theta_{S_i})^2$, where ${\theta}_{S_i}$ are the true underlying means for each observation. For non-normal data, we use the median absolute error (MAE) where $MAE=\sum_{i=1}^n |(\hat{\theta}(i)-\theta_{S_i}|$.

We obtain $4$ different estimates for the posterior mean of each of the $n$ observations as follows. The first two posterior mean estimates are from the Bayesian MCMC method implemented in the \textit{bcp} package \citep{erdman08} and the circular binary segmentation method from the \textit{DNAcopy} package \citep{olshen04,venkatraman07} in \textit{R}, both at the default parameters. 

We also use \textit{postcp}, with initial change-points from the estimated change-point set found by \textit{DNAcopy}, to estimate posterior probabilities of being in state $k$ for each observation $i$, $P(S_i=k|X_{1:n}=x_{1:n},S\in \mathcal{M}_K)$, and segment means $\hat{\theta}_k;k=1,\ldots,\hat{K}$. The posterior mean of observation $i$ is: 
\[
\hat{\theta}(i)=\sum_{k=1}^{\hat{K}} P(S_i=k|X_{1:n}=x_{1:n},S\in \mathcal{M}_K;\hat{\theta}_k) \hat{\theta}_k.
\]

To obtain another initial set of change-points for \textit{postCP}, we use another greedy procedure for quick model selection as follows. First, we use the greedy least-squares minimizing algorithm for a fixed $K$ number of segments to obtain an initial set of change-points. We then use the Viterbi algorithm to get the most probable set of change-points based on this initial set, and obtain the corresponding maximum likelihood estimates (MLEs) for the model parameters. The estimated number of segments $\hat{K}$ is that which maximizes the Bayesian information criterion (BIC) \citep{schwarz78,picard05}, \[ BIC(K)=LL-K^\prime \log(n),\] with respect to the number of segments, where $LL$ is the log-likelihood of the data, and $K^\prime$ is the number of parameters in the model. Posterior quantities are as previously described. 

We simulated a sequence of $n=500$ observations with $6$ change-points after $i=(22, 65, 108, 219, 252, 435)$. For normally distributed data, odd segments had mean $\theta_0=0.0$ and even segments had various values of $\theta_1$ with standard deviation $1.0$. For Poisson data, odd segments had mean $\theta_0=1.0$. We implement these procedures on $1000$ sets each of normal and Poisson distributed data for these $4$ methods. Results in Table \ref{MSE} report the MSE for normal data and MAE for Poisson data for different parameter values of $\theta_1$.

\begin{table}
\begin{center}
\caption{\label{MSE} Average error of different change-point methods} 
\setlength{\tabcolsep}{1mm}
\begin{tabular}{ll@{\hskip 1.0cm}cccc@{\hskip 1.0cm}cccc}\hline\hline
&&\multicolumn{4}{c}{$n=500$}&\multicolumn{4}{c}{$n=10,000$}\\\hline
$\theta_0$&$\theta_1$&cbs&cbs&greedy&bcp&cbs&cbs&greedy&bcp\\
&&&+postCP&+postCP&&&+postCP&+postCP&\\\hline
\multicolumn{10}{c}{normal distribution}\\
0.0&0.25&0.017&0.017&0.017&0.016&0.014&0.013&0.014&0.013\\
   &0.50&0.058&0.055&0.054&0.045&0.021&0.018&0.025&0.015\\
   &0.75&0.083&0.074&0.080&0.051&0.020&0.016&0.025&0.015\\
   &1.00&0.068&0.055&0.074&0.052&0.018&0.014&0.022&0.015\\
   &1.25&0.055&0.043&0.053&0.051&0.018&0.014&0.022&0.015\\
   &1.50&0.050&0.039&0.043&0.047&0.017&0.013&0.021&0.014\\
   &1.75&0.049&0.039&0.040&0.045&0.017&0.013&0.020&0.015\\
   &2.00&0.047&0.037&0.037&0.043&0.015&0.012&0.019&0.014\\
   &2.25&0.045&0.036&0.036&0.041&0.015&0.012&0.018&0.015\\
   &2.50&0.042&0.034&0.034&0.039&0.014&0.011&0.017&0.017\\\hline
   \multicolumn{10}{c}{Poisson distribution}\\
1.0&2.0&0.225&0.225&0.154&0.188&0.0438&0.045&0.051&0.062\\
   &3.0&0.112&0.112&0.119&0.196&0.0451&0.045&0.050&0.057\\
   &4.0&0.114&0.116&0.122&0.178&0.0469&0.047&0.050&0.055\\
   &5.0&0.122&0.123&0.126&0.175&0.0477&0.046&0.051&0.057\\
   &6.0&0.123&0.123&0.126&0.171&0.0519&0.051&0.052&0.066\\
   &7.0&0.136&0.137&0.138&0.180&0.0529&0.053&0.055&0.080\\
   &8.0&0.137&0.136&0.138&0.175&0.0497&0.049&0.053&0.091\\
   &9.0&0.139&0.140&0.140&0.181&0.0540&0.054&0.056&0.114\\
  &10.0&0.149&0.149&0.148&0.187&0.0546&0.055&0.058&0.150\\
  &11.0&0.154&0.154&0.153&0.194&0.0548&0.055&0.055&0.175\\
\hline
\end{tabular}
\flushleft{Error of posterior means for different change-point uncertainty methods, in terms of Mean Square Error (MSE) for normally distributed observations and Median Absolute Error (MAE) for Poisson distributed observations.}
\end{center}
\end{table}

For normally distributed data and $n=500$, \textit{bcp} had lower MSE than the \textit{postcp} estimates when the intersegmental differences, or differences between the even and odd segments, were less than 1.0 standard deviations. This may be due to the frequentist nature of \textit{postcp} using point estimates of change-point location to calculate MLEs, which may not have been accurate at lower intersegmental differences. However, at larger intersegmental differences, the methods relying on $\textit{postCP}$ actually had lower MSE than \textit{bcp}. In particular, for larger intersegmental differences, the Bayesian method at the default parameters of \textit{bcp} were overly conservative. 

\begin{figure*}
\begin{center}
\subfigure[$\theta_1-\theta_0=1.0$]{
\includegraphics[scale=0.4]{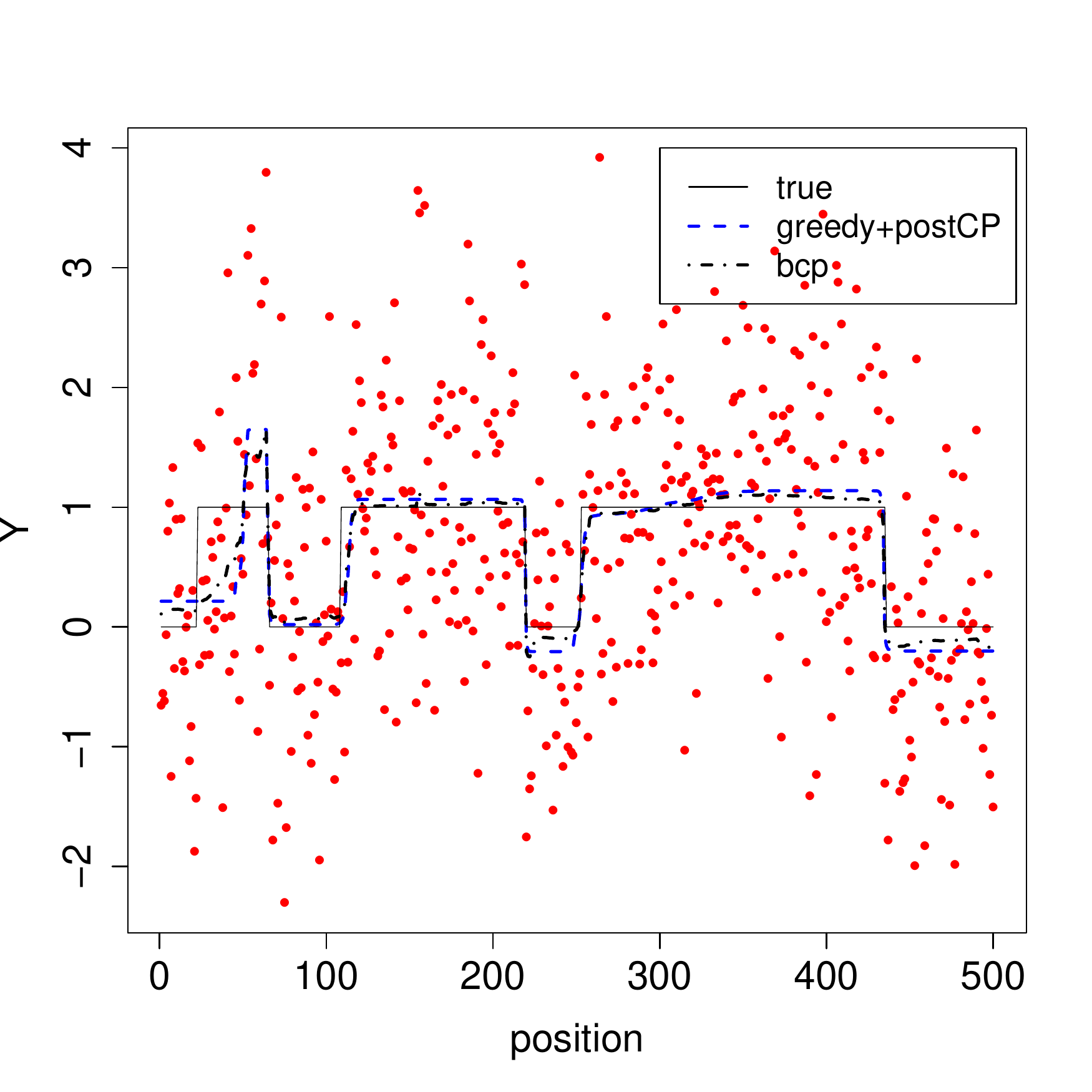}
\label{normsim1}} \\      
\subfigure[$\theta_1-\theta_0=2.0$]{
\includegraphics[scale=0.4]{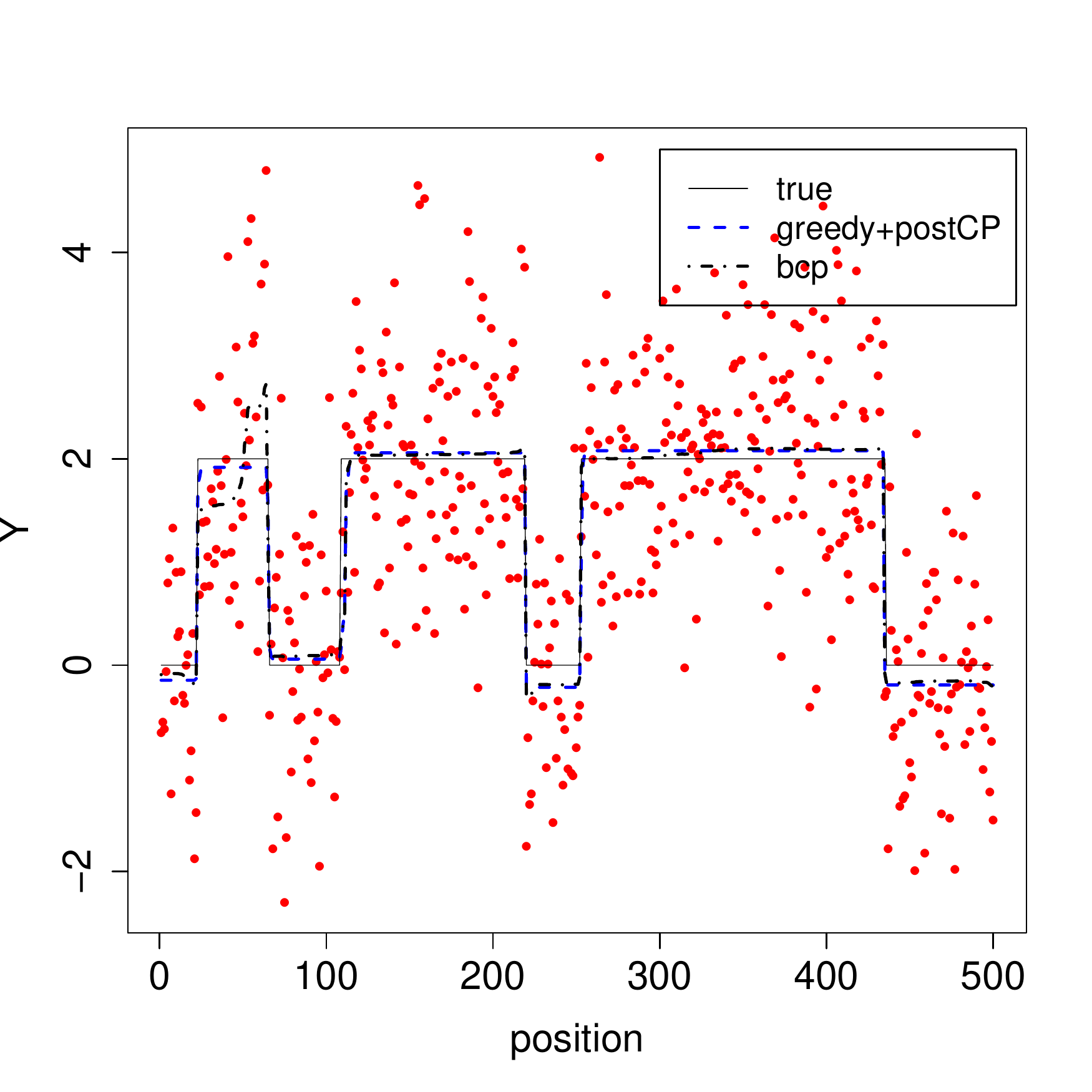}
\label{normsim2}}    
\caption{Plot of sample data $n=500$ from normal distribution in Table~\ref{MSE}. Underlying means in thin solid black lines, posterior means estimated by \textit{greedy+cp} in dashed blue lines and by \textit{bcp} in thick dash-dotted black lines. (a) Intersegmental differences of 1.0 SD, MSE for \textit{bcp} 0.043, for \textit{greedy+postcp} 0.066. The first change-point is not located precisely (at $i=50$) by the greedy method, with an extra change-point at $i=299$ from BIC. \textit{bcp} provided more conservative estimates slightly closer to true values in these regions. (b) Intersegmental differences of 2.0 SD, MSE for \textit{bcp} was 0.053, for \textit{greedy+postcp} was 0.040. More precise \textit{greedy+postcp} estimates after correct first change-point was entered.}
\end{center}
\end{figure*}

For illustrative purposes, Figures~\ref{normsim1} and \ref{normsim2} compare the true values of the underlying means to those found from \textit{greedy+postCP} and from \textit{bcp} in one specific set of generated data. For true intersegmental differences of $\Delta=\theta_1-\theta_0,=1.0$, $bcp$ had slightly lower MSE than \textit{postCP}. Inaccurate locations from the greedy algorithm and BIC resulted in a greater amount of error from \textit{postCP} compared to the relatively conservative posterior change-point probabilities from \textit{bcp}. However, at $\Delta=2.0$, \textit{postCP} performed slightly better than $bcp$, as the greedy algorithm and BIC provided the correct number of segments and accurate initial change-point estimates to $\textit{postCP}$. 

\begin{figure*}
\begin{center}
\subfigure[$\theta_1-\theta_0=2.0$]{
\includegraphics[scale=0.4]{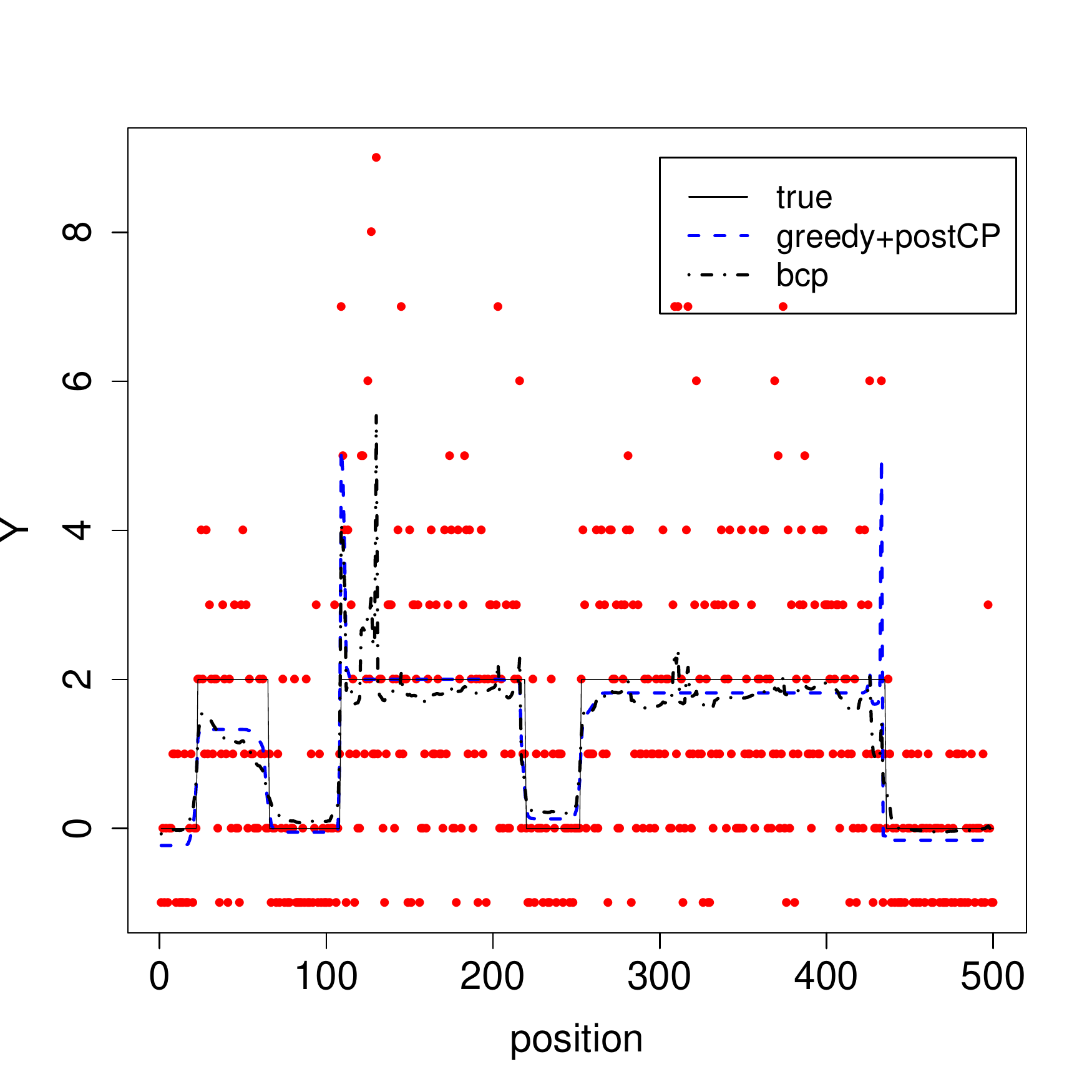}
\label{poissim1}}   \\    
\subfigure[$\theta_1-\theta_0=4.0$]{
\includegraphics[scale=0.4]{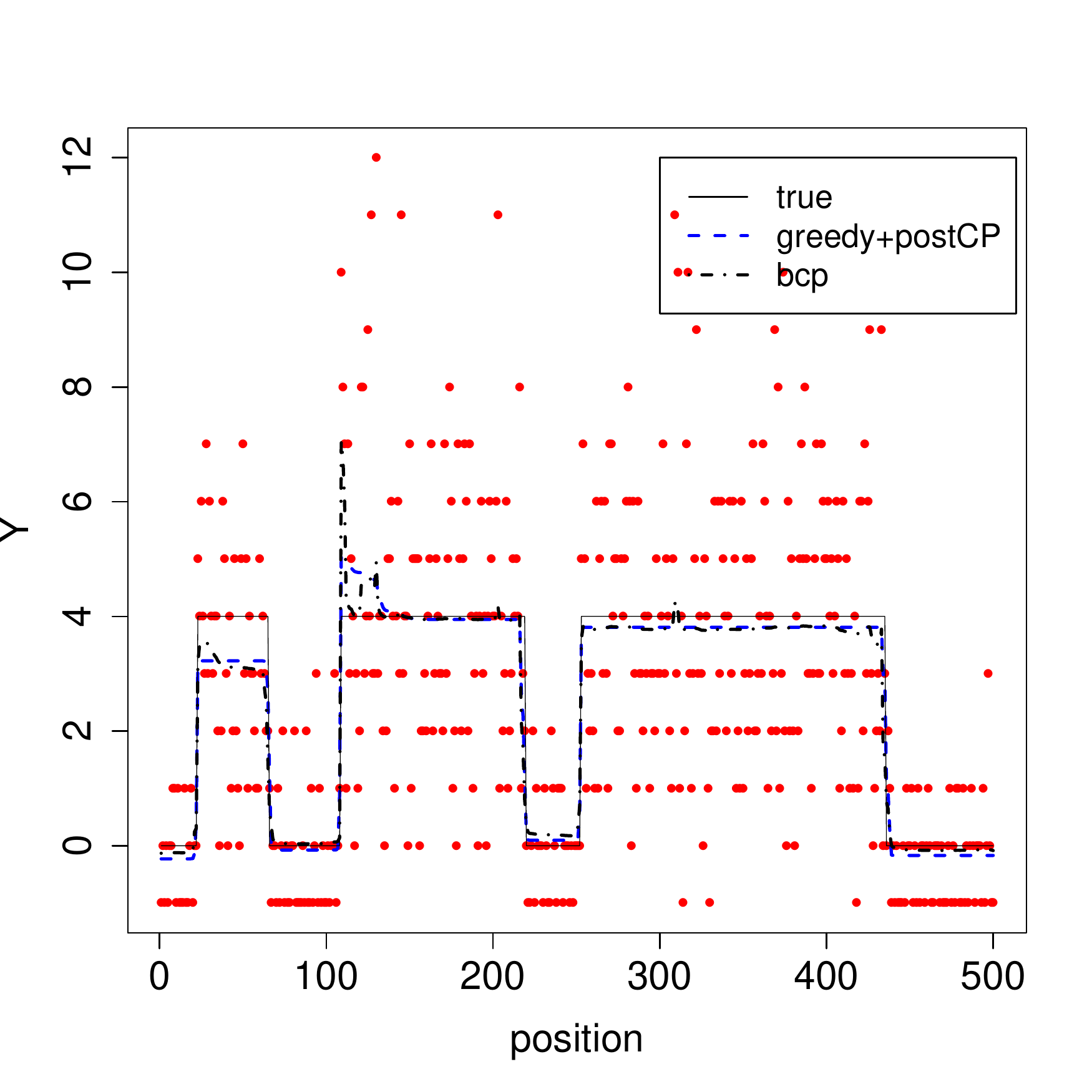}
\label{poissim2}}    
\caption{Plot of sample data $n=500$ from the Poisson distribution used in Table~\ref{MSE}. Underlying means in thin solid black lines, posterior means estimated by \textit{greedy+cp} in dashed blue lines and by \textit{bcp} in thick dash-dotted black lines. In both situations, the Bayesian estimates from \textit{bcp} provided irregularly shaped posterior mean curves. (a) $\theta_0=1$ and $\theta_1=3$, median absolute error for \textit{bcp} was 0.205, for \textit{greedy+postcp} was 0.172.  (b) $\theta_0=1$ and $\theta_1=5$, MSE for \textit{bcp} was 0.180, for \textit{greedy+postcp} was 0.175.}
\end{center}
\end{figure*}

For Poisson distributed data, the \textit{postCP} based methods had lower MAE than \textit{bcp}, showing the flexibility of the \textit{postCP} procedure which can adapt to non-normal data. 

Figures~\ref{poissim1} and ~\ref{poissim2} compare the true values of the underlying means for one set of Poisson data, for intersegmental differences of $2.0$ and $4.0$, respectively. It is evident that $postCP$ provides more appropriate posterior mean estimates as $bcp$ is not adapted for non-normally distributed data, with highly irregularly-shaped curves for the posterior means.

We also repeat the procedure on a larger sequence of $10,000$ observations, with $39$ change-points randomly selected from a uniform distribution within these observations, such that no segment is less than $25$ observations long. While there is less deviation in the posterior means due to larger segment sizes, the problem of estimating the number and location of change-points is made more difficult. For normal data using \textit{postCP} after initialization by \textit{cbs} showed similar patterns to those from $n=500$, obtaining lower MSE than $bcp$ other than for very small intersegmental differences. However, using \textit{postCP} after the greedy algorithm and BIC did not fare as well. Due to the larger amount of change-points, a more effective change-point location estimator than greedy least squares is recommended in these situations.

The exact \textit{postCP} algorithms also provided some computational advantages in terms of model selection over the Bayesian MCMC method \textit{bcp}, which also has linear complexity. For $n=500$, the entire model selection procedure averaged 0.07 seconds with the greedy+postcp model selection procedure, 0.09 seconds for cbs+postcp and 0.9 seconds for \textit{bcp}. For $n=10,000$, the respective runtimes were 1.37 seconds for cbs+postcp together and 15.7 seconds for \textit{bcp}.

\section{Detecting copy number variation in SNP array colorectal cancer data\label{CNV}}

To illustrate the efficiency of the constrained HMM model, we apply the methods to a copy number variation profile obtained through current SNP (single nucleotide polymorphism) microarray technology \citep{staaf08}. The data contain $261,563$ SNPs across the genome obtained from an Affymetrix chip from a colorectal cancer tumor cell line  \citep{deroock10} with intensities providing information regarding mutations, specifically duplications and deletions. After normalization, the log of the intensities in the tumor sample were compared to normal values, obtained from a reference sample, to obtain the log-reference ratios (LRR) across each of $23$ chromosomes. 

\begin{figure*}
\begin{center}
  \includegraphics[scale=0.5]{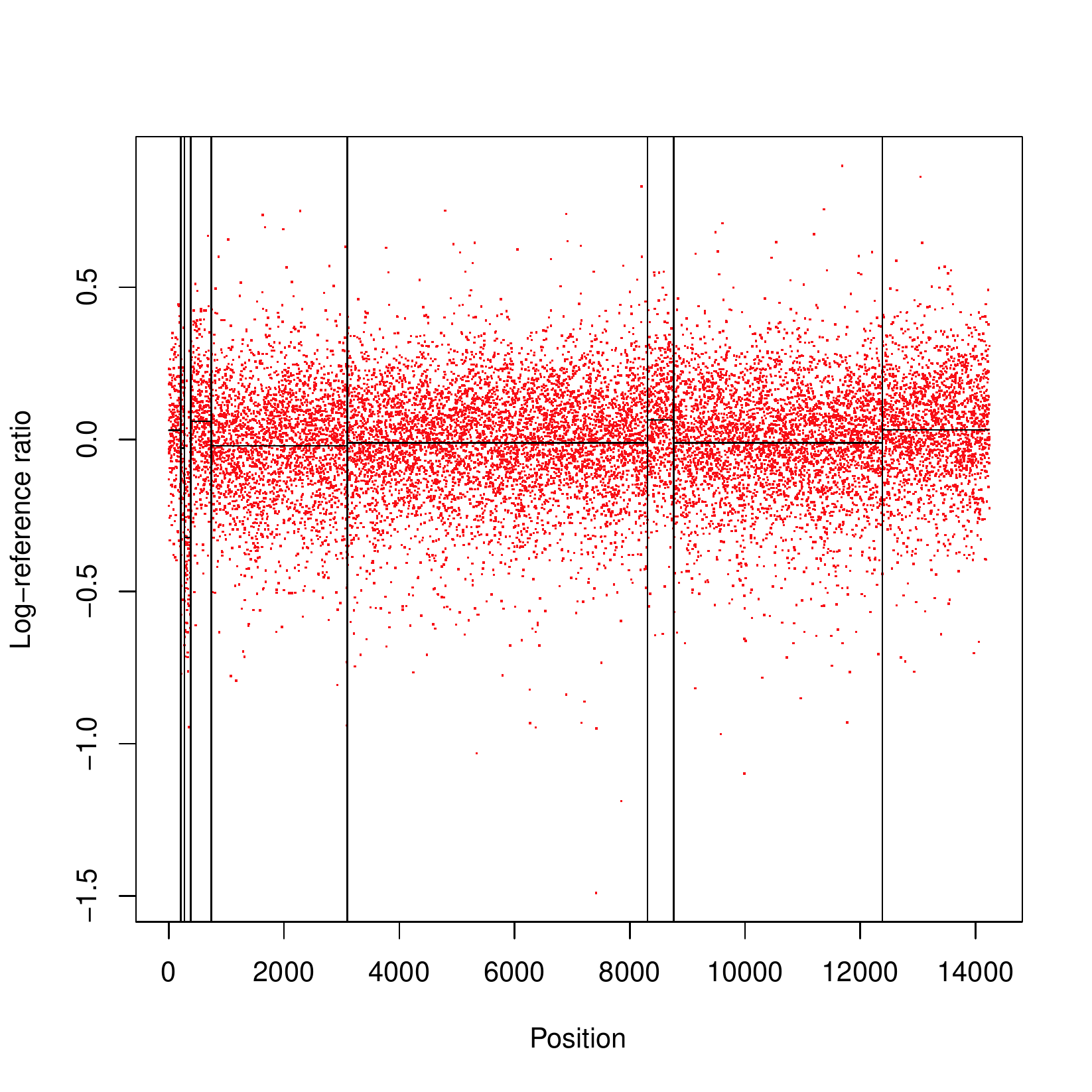}
\caption{Plot of 14,241 log-reference ratios (LRR) for chromosome~10. Horizontal lines are the estimated means within segments, vertical lines are the ten change-points estimated by DNAcopy.}
\label{best_seg_chr10}       
\end{center}
\end{figure*}

We use the circular binary segmentation (CBS) procedure implemented in a widely used segmentation package, DNAcopy \citep{olshen04,venkatraman07} for the statistics software \textit{R}. This package obtains an initial estimate of change-points in LRR, for each of the 23 chromosomes in tumor sample 103. We ran DNAcopy, after smoothing, at the default parameters of $\alpha=0.01$.  Table~\ref{seg_chr_10} displays the best segmentation found within for $14,241$ SNPs from chromosome~10. This chromosome was selected since both small (second) and large (tenth) segments were found, with adjacent segment differences ranging from 0.04 to 0.58 (Figure~\ref{best_seg_chr10}). 

\begin{table}
\begin{center}
\caption{\label{seg_chr_10} Segmentation from CBS for chromosome~10} 
\setlength{\tabcolsep}{1mm}
\begin{tabular}{lrrrr}\hline\hline\noalign{\smallskip}
Segment&Start&End&Size&Mean\\\hline
 1&    1&  211& 211& 0.031\\
 2&  212&  215&   4&-0.552\\
 3&  216&  273&  58&-0.028\\
 4&  274&  383& 110&-0.322\\
 5&  384&  736& 353& 0.060\\
 6&  737& 3091&2355&-0.021\\
 7& 3092& 3102&  11&-0.477\\
 8& 3103& 8308&5206&-0.011\\
 9& 8309& 8760& 452& 0.064\\
10& 8761&12383&3623&-0.011\\
11&12384&14241&1858& 0.031\\\hline
\end{tabular}
\flushleft{
Best segmentation results from circular binary segmentation algorithm \citep{olshen04}. Information from $K=11$ segments, with index of start and end of segment, size of the segment, and the mean within the segment. The pooled standard deviation across the segments was $\sigma=0.188$.}
\end{center}
\end{table}

We ran \textit{postCP} using the set of change-points identified by DNAcopy as initial segmentation to obtain estimates of the posterior probabilities of the change-points being at these ten locations. We assumed a homoscedastic normal model for the observations. The forward-backward algorithm was practically instantaneous (less than 0.1 seconds) for this sequence of over $14,000$ observations on a mid-range dual-core 2.5 GHz, 4GB RAM laptop PC. Not surprisingly, the most narrow confidence intervals, and most precise change-point estimates, were found for larger differences (Table~\ref{cp_chr_10}). Of the ten estimated change-points, six separated segments whose means differed by greater than one standard deviation; these points all were found to have posterior change-point probabilities greater than 0.5. Eight of the ten change-points from DNAcopy had the highest posterior change-point probabilities for their positions, with the exception of the fifth and tenth change-points whose probabilities were slightly lower than the respective maximum of their position.
\begin{table}
\begin{center}
\caption[Change-points from CBS for chromosome~10]{\label{cp_chr_10} } 
\setlength{\tabcolsep}{1mm}
\begin{tabular}{lrrr|rrrrr}\hline\hline\noalign{\smallskip}
Change&CP&Post&$\Delta$&\multicolumn{5}{c}{Size of confidence interval}\\
point&Est&Prob&Mean&ci:0.5&0.6&0.7&0.8&0.9\\\hline
 1&211&0.973&-0.582&  1&  1&  1&  1&  1\\
 2&215&0.918& 0.523&  1&  1&  1&  1&  1\\
 3&273&0.556&-0.293&  1&  2&  2&  2&  3\\
 4&383&0.580& 0.381&  1&  2&  2&  2&  3\\
 5&736&0.028$^a$&-0.081& 31& 37& 46& 53& 61\\  
 6&3091&0.860&-0.456&  1&  1&  1&  1&  2\\
 7&3102&0.880& 0.466&  1&  1&  1&  1&  2\\
 8&8308&0.050& 0.075& 21& 30& 50& 91&132\\
 9&8761&0.064&-0.075& 15& 21& 48& 63& 78\\
10&12383&0.006$^b$& 0.042&233&336&412&501&522\\\hline
\end{tabular}
\flushleft{
Information for 10 change-points identified within chromosome~10. The posterior probability of a change was at least 0.50 for all change-points with a moderate change ($>\sigma=0.188$), while wider confidence intervals were found when segment differences were smaller. $^a$: point 721 had slightly higher change-point probability (0.031),$^b$: point 11943 had slightly higher change-point probability (0.008).}
\end{center}
\end{table}

\begin{figure*}
\begin{center}
\subfigure[Change-points 1 to 5]{
\includegraphics[scale=0.35]{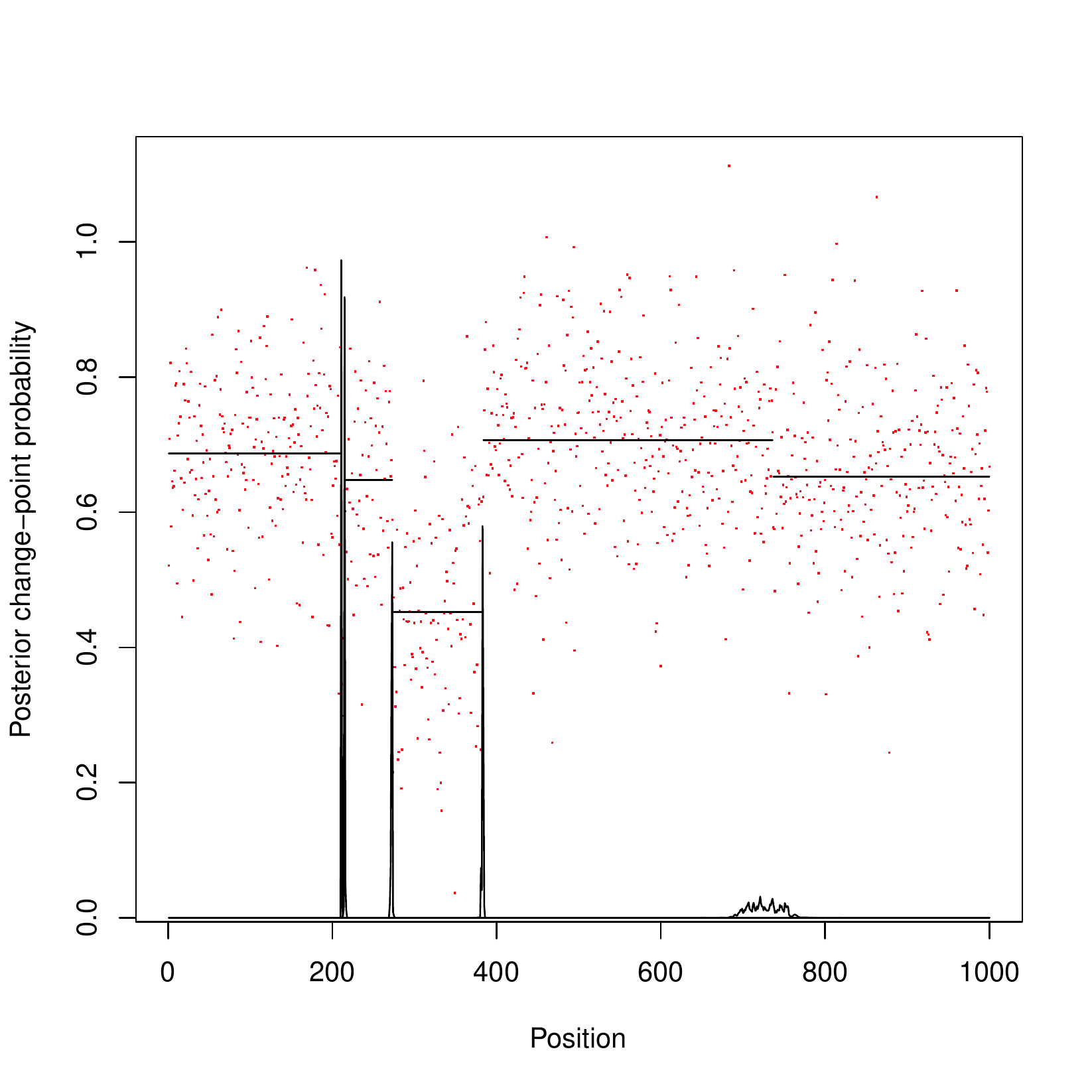}
\label{cp_chr10_15}}       
\subfigure[Change-point 10]{
\includegraphics[scale=0.35]{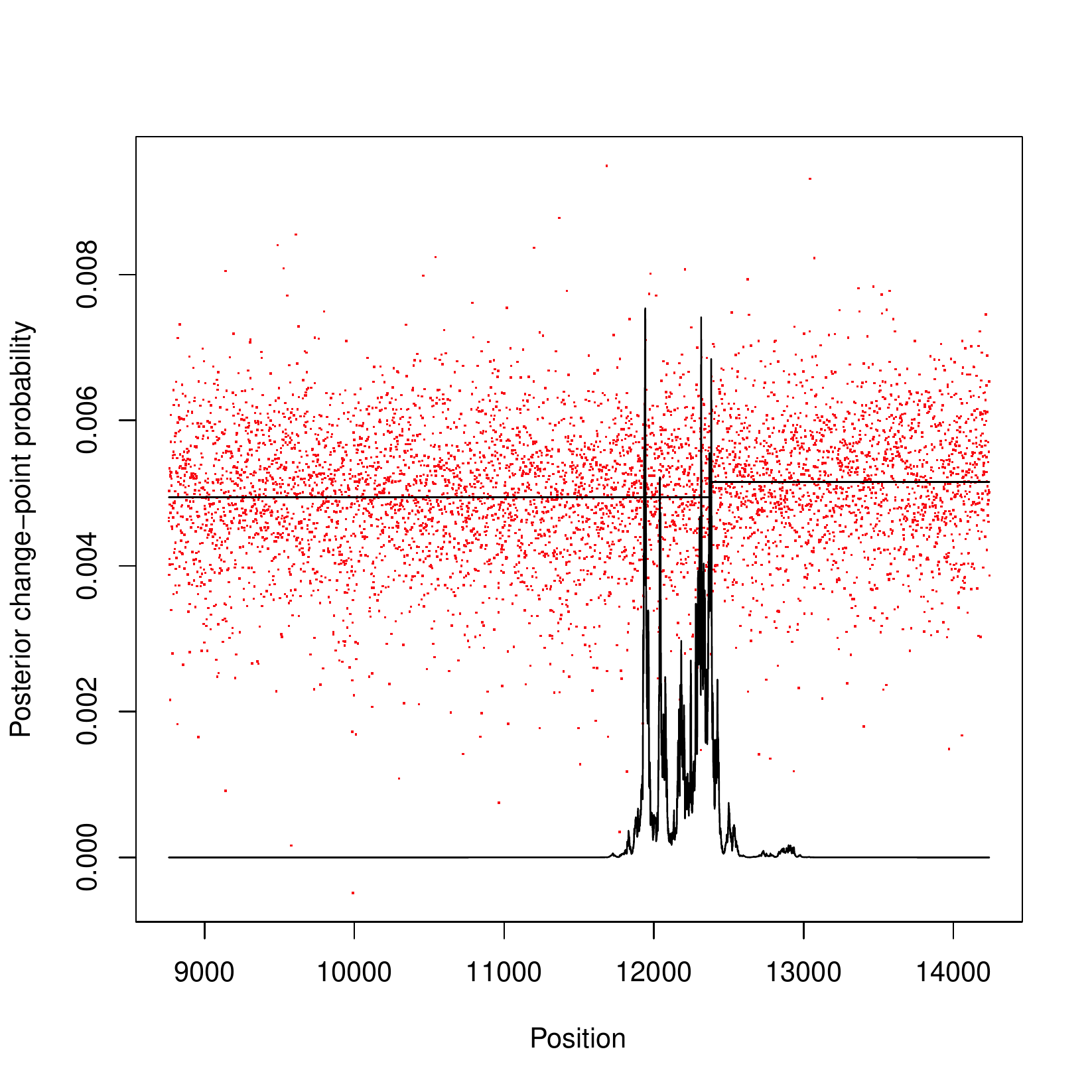}
\label{cp_chr10_10}}    
\caption{Plot of estimated posterior change-point probabilities. Dots are LRR data with horizontal lines the estimated means within segments scaled to the probability axes. (a) For first five change-points: The first four change-point estimates (SNPs at position: $211,215,273,383$) are precise, however there is a wide uncertainty around the fifth change-point (position: 736). (b) For tenth change-point: There is a very wide confidence interval for this change-point estimate. The change-point estimate from the CBS algorithm (position 12,383) is the fourth rightmost peak.}
\end{center}
\end{figure*}

Figure~\ref{cp_chr10_15} displays the posterior change-point probabilities for the first five change-points. Given the much narrower segment lengths, and greater segment differences, the probabilities are much higher for the first four change-points than the fifth change-point, whose 90\% confidence interval was greater than 60 SNPs wide. Figure~\ref{cp_chr10_10} displays the posterior change-point probabilities for the tenth change-point, whose 90\% confidence interval was greater than 500 SNPs wide. Note that the shapes of the change-point distributions are also highly irregular.

\section{Discussion\label{discussion}}

A common point of interest in current genomics studies is to find genetic mutations pointing to phenotypes susceptible to diseases \citep{redon06} such as cancers or Type II diabetes. The use of change-point analysis for detecting copy number variation (CNV) is a critical step in the characterization of DNA, including tumoral genes associated with cancer. A CNV may aid in locating a genetic mutation such as a duplication or deletion in a cancerous cell that is a target for treatment. However, due to the often large nature of genomics data sequences it is often difficult to reliably identify CNVs. The latest technologies for detecting CNV, such as single nucleotide polymorphism (SNP) arrays, are able to produce sequences of tens or hundreds of thousands of observations. 

This paper describes a procedure that extracts further useful information from segmenting large sequences such as those found in CNV detection in SNP array data. The unsupervised HMM, which is a level-based approach used in many current implementations, includes forward-backward calculations in linear time. However they are less flexible than segment-based methods, in part because they require prior transition probabilities between states that assume a geometric prior on segment lengths, which may be inappropriate in some situations. On the other hand, most segment-based methods can allow for a wider variety of specified priors, but in many situations are unfeasible due to their quadratic complexity for large data sets such as those generated by SNP arrays. 

The described constrained HMM model's complexity of $O(Kn)$ rather than $O(K n^2)$ enables it to handle very large data sets in reasonable time. The procedure allows for the estimation of the full joint distribution of change-points conditional to the observations, allowing for practical estimation of confidence intervals. In high-resolution data sets with change-points to be unlikely detected at the exact correct position, the confidence intervals may yield important information. In SNP array technology, the change-points need to be detected with high-precision and the differences between segments may not be very large. Additionally, overlapping confidence intervals across several different cell tumors or lines may identify associated copy number variations, and help in identifying similar disease phenotypes and treatments in patient subgroups. 

The emission distribution $g_{\theta}(x)$ of the observed data can also be easily specified in this constrained HMM procedure, unlike other implementations specifically developed for normally distributed large-scale data. The ability of \textit{postCP} to easily adapt to a wide variety of distributions through the specification of the emission distribution is especially important with next-generation sequencing platforms often producing data that closely follows the Poisson or Negative Binomial distributions \citep{shen11}. 

The described methods are a useful tool in the segmentation of large-scale sequences such as those involving CNVs, specifically when combined with any current implementation designed for detecting an ideal set of change-points. For example, a fast dynamic programming algorithm of order $O(n \log n)$ developed by \citet{comte04} obtaining an initial segmentation of the points is to be bundled in the \textit{postCP} package. While the constrained HMM approach is sensitive to its starting point and tends to provide a local rather than global optimum if the initial starting points are misspecified, simulations in this paper showed that in practice, \textit{postCP} provided estimates of posterior means that were comparable to, or sometimes better than, a currently implemented Bayesian MCMC method. This occurred in situations in which the Bayesian method was specifically designed, that is for assessing multiple change-points in normally distributed data.  

A planned extension of our method is to combine the segment-based and level-based approaches by limiting possible values of segment parameters $(\theta_1,\ldots,\theta_K)$ to $L\leq K$ different values. Further practical applications include more complex models to simultaneously model multiple datasets, or accounting for values from multiple patients and samples as random effects in a mixed model. Another practical extension is to handle multi-dimensional output, such as current technology for copy numbers \citep{staaf08} that simultaneously includes LRR and baseline allelic frequency (BAF) data.

\section*{Acknowledgements}
We would like to thank Dr. Pierre Laurent-Puig of the INSERM S-775 biomedical laboratory of Paris Descartes for permitting us to use the colorectal cancer SNP array data used in our illustrative example on high-resolution data, and Mr. Mael Th\'{e}paut for providing the normalized data set. We would also like to thank Prof. St\'{e}phane Robin and Dr. Guillem Rigaill for providing us with the breast cancer data from cell line BT474 they used in the previous publication.

\bibliographystyle{elsarticle-harv}      


\bibliography{postCP}

\end{document}